\documentclass[final,5p,times,twocolumn]{elsarticle}

\usepackage{graphicx}

\usepackage{amssymb,amsmath,amsthm}

\usepackage{url}

\journal{Reliability Engineering and System Safety}

\begin{document}

\begin{frontmatter}



\title{Algorithm for Resource Redistribution \\ Required for Recovery of Society after Large Scale Disasters}


\author{Vasily Lubashevskiy, Taro Kanno, and Kazuo Furuta}

\address{Department of Systems Innovations, School of Engineering, University of Tokyo, 
7-3-1 Hongo Bunkyo-ku, 113-8656 Tokyo, Japan}

\begin{abstract}
The recovery of society after a large scale disaster generally consists of two phases, short- and long-term recoveries. The problem of short-term recovery is rather close to the problem of resilience in their goal, namely, bouncing the damaged system back to the operating standards. The present paper proposes an algorithm for the vital resource redistribution required for implementation of the short-term recovery. 
The developed model is based on the cooperative interaction of cities during the resource redistribution, ordering the cities according to their priority in resource delivery, and a generating a semi-optimal plan for the desired redistribution. Nonlinear effects caused  by the city limit capacities are taken into account. Two types of systems, ``uniform'' and ``centralized'', are studied numerically. In particular it is demonstrated that the cooperation effects  are able to shorten substantially the duration of the process required for its implementation. 

\end{abstract}


\end{frontmatter}


\section{Introduction}
\label{}

In recent years the problems of disaster mitigation and resilience have attracted much attention. As far as mitigation of large scale disasters is concerned, two phases, short- and long-term recoveries, can be distinguished. Use of these terms has a long history \cite{national1979comprehensive}, nonetheless, the appropriate classification of recovery phases is required especially for efficient emergency management of large scale disasters \cite{DHS2008,MalcolmBaird2010a,DHS2013}.   

Following the cited materials we consider the short-term recovery to be \textit{mainly} aimed at restoring the vital life-support system to the minimal operating standards. Generally this system comprises many individual components and the corresponding services, in particular, sheltering, feeding operations, emergency first aid, bulk distribution of emergency items, and collecting and providing information on victims to family members. One of the basic requirements imposed on the short-term recovery is the beginning of its implementation within the minimal time. For example, the aforementioned services have to start their operation within 8 hours according to the Disaster Recovery Plan of 
State of Illinois \cite{DRP_Illinois_July2012_STRecovery}.

To mitigate aftermath of a large scale disaster cooperation of many cities is required, because the amount of resources initially accumulated in an affected area can be insufficient to recover all the individual components of the vital life-support system. Thereby the implementation of the short-term recovery is directly related to an efficient resource redistribution. This redistribution cannot be predetermined because of the unpredictability of disaster consequences. It is possible only to formulate rather general requirements for this process. First, the supply to an affected area must start practically immediately in order to recover the life-support system. Second, the resource supply should be decentralized, otherwise, its centralized management can be a `bottleneck' that delays the responsive and adaptive delivery of resources or aid \cite{managprob:panda2012}.

The purpose of the present paper is developing an algorithm by which such resource redistribution can be implemented. It is worthwhile to note that this algorithm is not reduced to the basic logistic problems. The matter is that all the necessary resources cannot be sent out at one time. A city not affected by the disaster and involved in the resource supply requires some finite time for preparing a ``quantum'' of resources for transportation. Such cities operate in a step-by-step manner: ``prepare a quantum of resources--send it out--prepare a new one.''  As a result, the event of sending a resource quantum from a city $A$ to an affected city $B$ excludes the city $A$ from the process of resource redistribution for the time required to prepare the next ``quantum.'' This ``temporal'' exclusion makes the source located in the city $A$ unachievable for the other affected cities, which modifies the supplying network. Similar modifications are initiated continuously by all the cities in the affected area, which makes it inevitable to consider the supplying network dynamics. Nowadays, the problems related to dynamical or temporal networks is regarded as an individual branch of science not related to classical logistics \cite{holme2012temporal}.

In what follows we will construct an algorithm governing the desired resource redistribution and analyze numerically the dynamics of the supply process. Two particular cases of the initial resource distribution, centralized and uniform, will be studied in detail. Besides, the supply dynamics depending on the number of elements in the suppling network and the city limit capacity will be investigated. 

\section{Model}

\begin{figure}[t]
\begin{center}
\includegraphics[width=0.7\columnwidth]{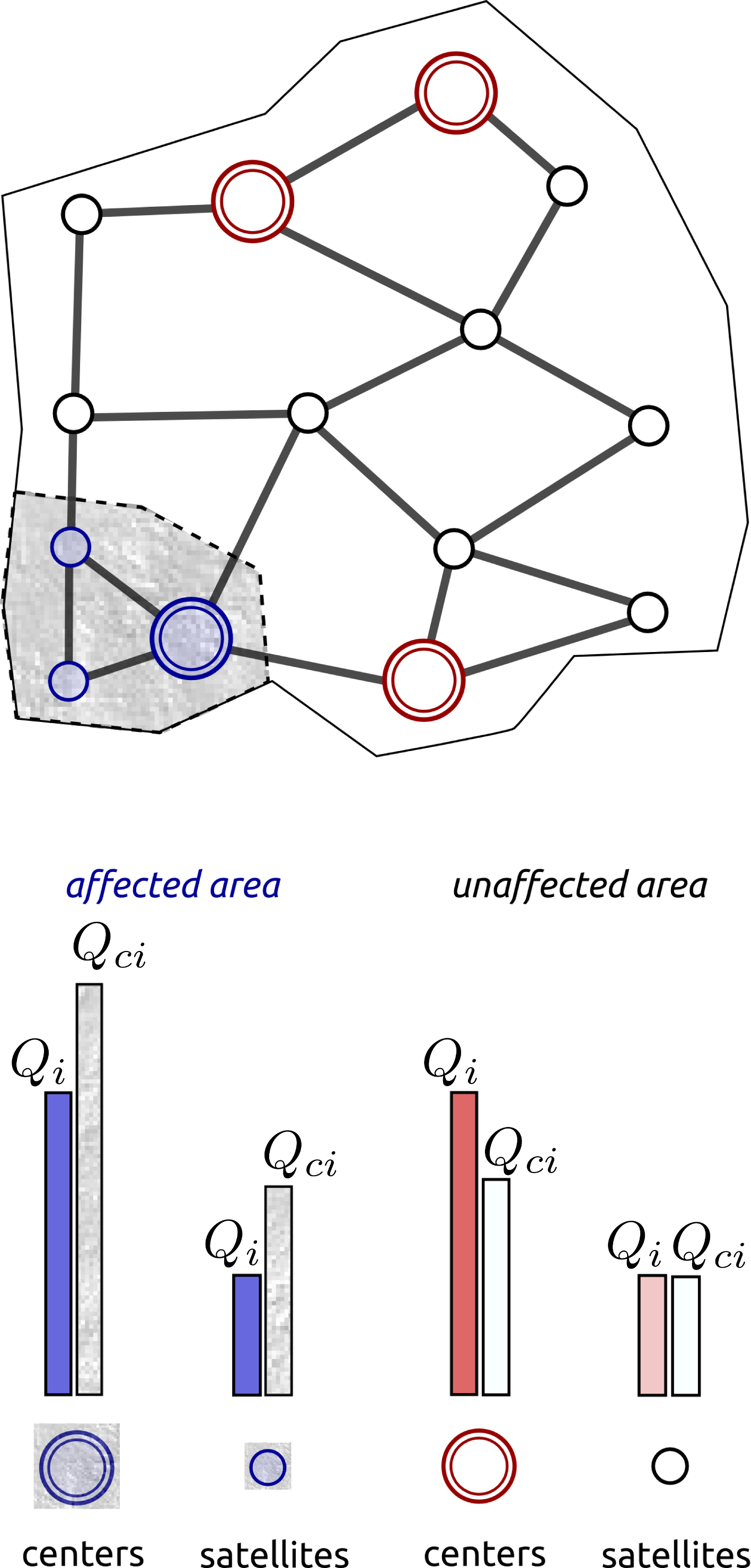}
\end{center}
\caption{A schematic illustration of the ``centralized'' system. The affected area is shadowed and for the cities there the minimal critical level $Q_{ci}$ becomes higher than the current amount of resources $Q_i$.}
\label{F1}
\end{figure}

\subsection{Model background}\label{modback}

The Great East Japan Earthquake occurred along the eastern coast of Japan on 11 March 2011 exemplifies large scale destructive disasters that necessitate the cooperation of many cities and even regions in mitigating the aftermath. The hypo-central region of this earthquake comprised several offshore prefectures (Iwate, Miyagi, Fukushima, and Ibaraki Prefectures) and have  ruptured the zone with a length of 500~km and a width of 200~km \cite{wiki:GEJ2011}. The terrible aftermath of the disaster initiated evacuation from some areas of these prefectures, thousands houses were destroyed, many victims required medical assistance. Obviously none of the affected cities was able to recover only by its local resources, practically all the non-affected cities in these prefectures were involved into this process. 
New shelters were urgently created in many non-affected regions, some highways were closed for private vehicles, flows of required pure water, food, medical drugs, fuel, etc. was redirected to the damaged cities. 
The ability to modify urgently the supplying system is one of the crucial points for a high resilience of the system as a whole.
These Japanese prefectures can be one of the best examples of the system, which overcame the disaster and recovered to its normal state. In numerical simulation to be described below some of the system parameters were evaluated using, for example, the real data for Fukushima prefecture. Namely, the total number of residents is evaluated as $10^6$, the area of the region treated as a certain administrative unit responsible for mitigating the aftermath is set about $10^4$~km$^2$, the mean distance between the neighboring cities in this region is 40--50~km, as a results, the number of cities that can be involved into recovering the affected region may be about 5--50.

\subsection{System under consideration}

The system is modeled as a collection of cities connected with one another by a transport network. Initially in each city $i$ there is some amount of resources $Q_i$ depending on the number $N_i$ of residents in it. Under the normal conditions this amount of resources is excessive and substantially exceeds the minimal critical level $Q_{ci}$ required for its residents to survive during a certain period of time, $Q_i > Q_{ci}$. Naturally the magnitude of the quantity $Q_{ci}$ depends on the number $N_i$ of residents in a given city $i$; the larger the number of residents, the higher the required level of resources $Q_{ci}$. One of the consequences of a large scale disaster is the increased demand for the vital resources in the affected cities. This is modeled as the essential increase in the corresponding magnitudes of $Q_{ci}$ and the opposite inequality $Q_i < Q_{ci}$ holds for the affected cities. Naturally the inequality
\begin{equation}\label{in:1}
\sum_i Q_i > \sum_i Q_{ci}
\end{equation}
must hold still after the disaster. Actually inequality (\ref{in:1}) is the mathematical implementation of the requirement that the given system is capable to survive as a whole during a certain time without external help.

To examine the dynamics of supply process two limit cases of the initial resource distribution will be modeled. The first one is a ``uniform'' system. In this case all the cities are supposed to be initially equal in all the parameters. The second one is a ``centralized'' system which comprises a collection of small cities (``satellites'') and ``centers''.  In the ``satellites" the number of residents is less than in the ``centers" and for them the equality $Q^{satelit}_i=Q^{satelit}_{ci}$ is assumed to hold at the initial stage. To make the systems comparable the integral parameters 
\begin{equation}\label{Eq.resource balance}
\sum_i Q^\text{centalized}_i = \sum_i Q^\text{uniform}_i
\end{equation}
\begin{equation}\label{Eq.residents balance}
\sum_i N^\text{centalized}_i = \sum_i N^\text{uniform}_i
\end{equation}
are supposed to be equal.

An example of the ``centralized'' system is illustrated schematically in Fig.~\ref{F1}. It depicts a collection of cities linked to one another with a transportation network. Its part affected by the disaster is shown as a shadowed region.

\section{Resource redistribution algorithm}

This section presents the logic of mechanism and its realization algorithm. Leaping ahead, we note that there is a essential difference between the problem under discussion and problems of classical logistics. The matter is that in our case the network is dynamical. All the resources are located in some warehouses in or near the cities and their capacities are limited with respect to the amount of resources as well as the operation ability (limited number of loading vehicles). In particular, when the first group of parcels in one of the warehouses are sent, it takes a time to prepare another one for sending. During this interval the resources in the given warehouse are not accessible for all the other cities and this warehouse became temporally ``cut off" from the network. Such behavior of the system endows the process with nonlinearity. To take into account this effect the algorithm uses the time distances between cities instead of geographic ones and their specific values depend on the intensity of supply flow.

At the initial step all the cities that are accessible provide the information about their state, namely, the available amount of resources $Q_i$, the minimal critical amount $Q_{ci}$ required for their individual surviving, and the number $N_i$ of citizens. The characteristics of the transportation network are assumed to be also given, it is the matrix $\mathbf{D}=\|d_{ij}\|$ whose element, e.g., $d_{ij}$ specifies the minimal time distance between city $i$ and $j$. To describe the states of cities let us introduce the value
\begin{equation}\label{eq:theta}
	\theta_i = \frac{Q_i - Q_{ci}}{Q_{ci}}\,.
\end{equation}
When the information about a given city $i$ is not available, the corresponding value is set, by definition, zero, $\theta_{i}=0$. If $\theta_i < 0$ then its magnitude quantifies the lack of vital resources in relative units. The quantity $S_i = \theta_i N_i$ is actually the number of people being under the level of surviving. It will be used in specifying the priority of the cities in the resource redistribution queue.  
To avoid the discussion about ethics or morality of the priority choice we appeal to the following historical example.  Baron Dominique Jean Larrey, surgeon-in-chief to Napoleon's Imperial Guard, articulated one of the first triage rule in 1792: ``Those who are dangerously wounded should receive the first attention, without regard to rank or distinction. They who are injured in a less degree may wait until their brethren in arms, who are badly mutilated, have been operated on and dressed, otherwise the latter would not survive many hours; rarely, until the succeeding day" \cite{Napaleon}. The minimal value of $S$ corresponds to the maximal number of residents which are not supplied with the vital resources and it endows us to mark that city as a most ``dangerously wounded''.

Because the main goal of resource redistribution just after the disaster is mitigation of consequences and minimization of the amount of victims, Table~\ref{T1} determines the priority of the resource redistribution.
\begin{table}[h]
\caption{The order of cities according to the resource redistribution priority. Here $M$ is the total number of cities in the given system.}\label{T1}
\begin{center}
\begin{tabular}{|c||c|c|c|c|c|}
\hline
$S_p$ & $S_1$ & $S_2$ & $\ldots$ & $S_{M-1}$ & $S_{M}$ \\
\hline
$p$ &   $1$ &  $2$ &  $\ldots$ &    $M-1$ &  $M$  \\
\hline 
\end{tabular}
\end{center}
\end{table}
\noindent
The order used in Table~\ref{T1} matches the inequality
\begin{equation}\label{eq:order}
S_1 \leq S_2 \leq \ldots\leq S_{M-1}\leq S_{M}
\end{equation}
and $i_1$, $i_2$, $\ldots$, are the corresponding indexes of these cities. 

In order to describe resource redistribution dynamics, let us introduce the following quantities. First, it is a certain quantum $h$ of resources that can be directed from a city to another. The second quantity is the time $\Delta t$ required for this quantum to be assembled for transportation. The third one  $c_i$ is the capacity of a given city $i$ specifying the maximal number of quanta which can be assembled during the time $\Delta t$. Introduction of these
quantities implies the realization of resource redistribution mainly via fast loading vehicles, for instance, tracks. In this case $h$ is the volume of resources transported by the typical vehicle individually, $\Delta t$ is the time necessary to load this vehicle, and $c_i$ is determined by the number of loading places and the capacity of loading facilities. 

The algorithm to be described below creates a complete plan of resource redistribution depending explicitly on the initial post-disaster system state. Namely, at the first step Table~\ref{T1} is formed using the initial data. The city $i_1$ is selected as the city with the wost situation. Then we choose a city $i_k$ such that
\begin{equation}\label{eq:dsearch}
d_{i_1 i_k} = \min_j d_{i_1 j}\quad \text{among} \quad Q_j - h \geq Q_{cj}\,.
\end{equation}
Then the prepared quantum is \textit{virtually} transported to city $i_1$ from city $i_k$ . It gives rise to the transformations
\begin{equation}\label{eq:transf}
\begin{split}
Q_{i_1} &\rightarrow Q_{i_1} + h\,,\\ 
Q_{i_k} &\rightarrow Q_{i_k} - h\,, \\ 
c_{i_k} & \rightarrow c_{i_k} - 1\,.
\end{split}
\end{equation}
The information about the given action is saved as a report of its virtual realization and comprises: ``city $i_k$ sent one quantum to city $i_1$ at time $t_\text{dep}$, the quantum is received at time $t_\text{arr}$''. These time moments are related via the equality
\begin{equation}\label{eq:tdeptarr}
t^\text{initial}_\text{arr} =  d_{i_1 i_k}\,.
\end{equation}
The superscript ``initial'' has been introduced to underline the fact that the initial value of the matrix $\mathbf{D}$ enters Exp.~\eqref{eq:tdeptarr}. Below it will be renormalized to take into account the nonlinear effects caused by the city limit capacity.

At the next step this procedure is reproduced again. Table~\ref{T1} is reconstructed, the logic of choosing the interacting cities is repeated with saving the relevant report. 

Since the maximal number of quanta that can be sent from a given city simultaneously is finite, there exist a situation where $c_j$ takes a zero value due to transformations~\eqref{eq:transf}. This effect is taken into account by renormalization of the matrix $\mathbf{D}$, which is a time distance matrix. Namely, when $c_j = 0$ we restore the initial value of $c_j$ and for all $i$
\begin{equation}\label{eq:decrD}
\begin{split}
d_{ij} &\rightarrow  d_{ij} + \Delta t\,,\\ 
t_{\text{dep},j} &\rightarrow t_{\text{dep},j}+ \Delta t\,.
\end{split}
\end{equation}
This procedure is terminated when at the next step 
\begin{equation}
\forall i: \quad S_i > 0\,.
\end{equation}
As a result, this algorithm generates the collection of reports which enables us to create a semi-optimal plan of resource redistribution for all the cities and the real process can be implemented.

\section{Results of Numerical Simulation}


\begin{figure}
\begin{center}
\includegraphics[width=0.5\columnwidth]{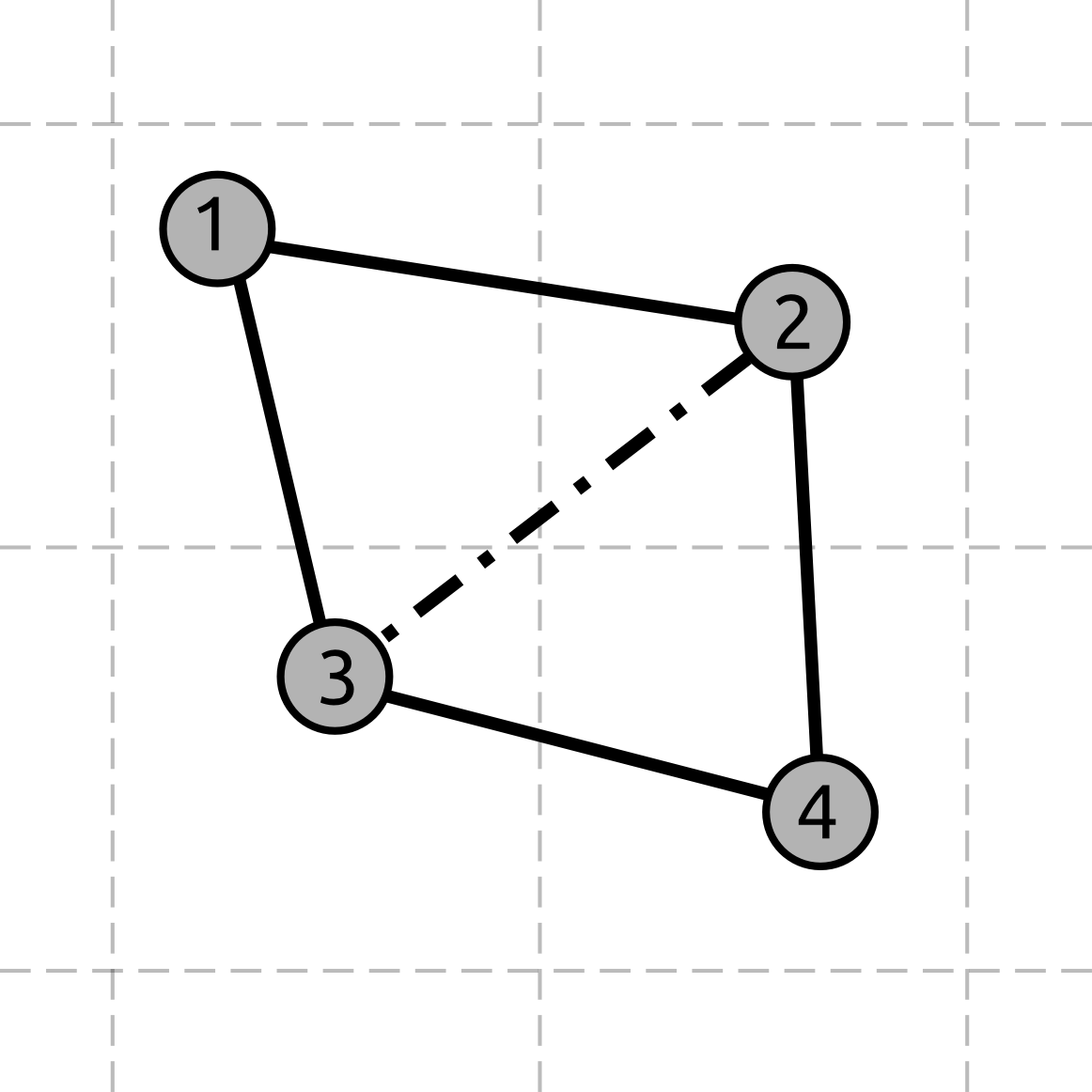}
\end{center}
\caption{Example of the city arrangement and the corresponding transportation network.}
\label{F:angle}
\end{figure}

\subsection{Details of modeling}

The purpose of the present section is to illustrate the features of the analyzed resource redistribution process. Keeping in mind the administrative units noted in Section~\ref{modback}, the following systems were studied numerically based on the developed model. Each of them is assumed to comprise 20 cities regarded as basic entities connected with one another by a transport network and the total population of these cities is set $P=2\times 10^6$. Two types of systems, ``uniform'' and ``centralized'' were analyzed individually. For specific purposes some system parameters, namely, the number of cities and the total number of residents were changed.  The amount of resources were measured in the units of resource quantum $h$, so we set $h=1$. To be specific the volume of one quantum is assumed to supply 100 residents with the critical amount of resources with some additional extra volume (60~\%) under the normal conditions. So the integral amount of resources initially allocated in the system is 
\begin{equation*}
\sum_i Q_i = \frac{P}{100}\,.
\end{equation*}
The mean time distance between the cities was varied from 40 to 120 minutes and the time $\Delta t$ necessary to prepare one resource quantum was set 5--15 minutes.

The transportation network was constructed in the following way. The region occupied by the given system is considered to be of a rectangular form and divided into $20$ (the number of cities) equal rectangles. Each rectangle contains one city placed randomly within it. At the first step the connections between the cities located in the neighboring rectangles are formed as illustrated in Fig.~\ref{F:angle}. For any arrangement of these four cities the "vertical" and "horizontal" connections are formed. A diagonal connection, for example, the connection 2-3 is formed if both of the opposite angles are less than $90^\text{o}$: $\angle{213}$ and $\angle{243}$ in Fig.~\ref{F:angle}. In this way we construct the matrix \textbf{D} of minimal time distances between the neighboring cities. The relationship between spatial and temporal scales was determined assuming the average speed of transporting vehicles equals 60 km/h. At the next step using Warshall's algorithm 
(see, e.g., \cite{discmathstr_kbr6})
%
%
we complete \textbf{D} to the matrix of the minimal time distances between any pair of cities. In the case of affected cities some of
the connections were cut up, however, without losing the graph connectivity.

\begin{figure}
\begin{center}
\includegraphics[width=0.9\columnwidth]{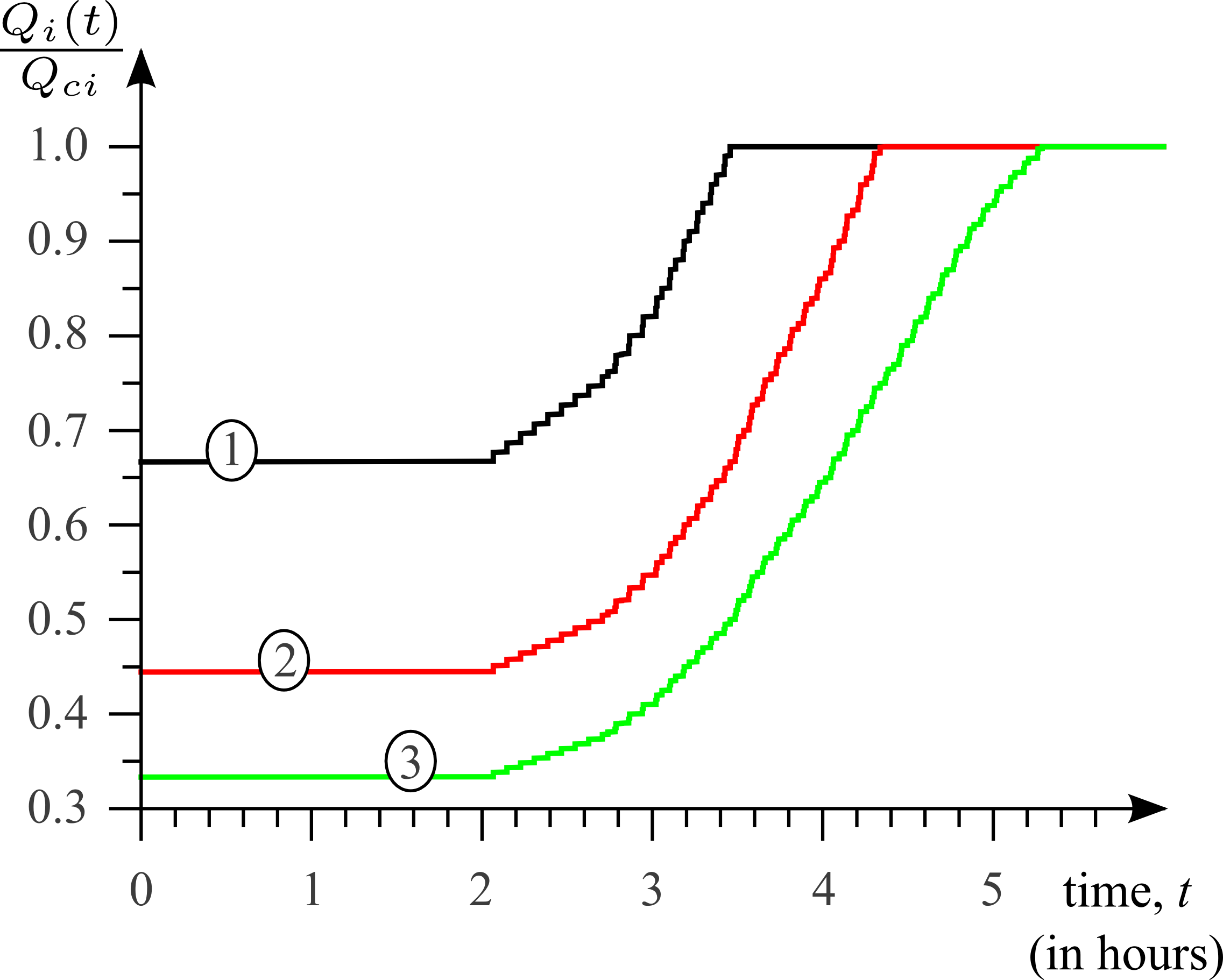}
\end{center}
\caption{The recovery dynamics of an affected city $i$. Curves 1, 2, 3 demonstrate the recovery dynamics when the degree of damage $a$ is equal to 1.5, 2.25, 3, respectively. All the other system parameters as well as the system topology were the same.}
\label{F:D123}
\end{figure}

\subsection{``Uniform'' system}

First, let us consider the results of numerical simulation for the ``uniform'' system. Three neighboring cities located at one of the rectangular corners were supposed to be damaged. Figure~\ref{F:D123} illustrates the dynamics of supplying an affected city $i$ according to the plan generated by the developed algorithm. Three curves represents the recovery dynamics of the affected city $i$ for three different degrees of damage, $a = \dfrac{Q^\text{affected}_{ci}}{Q_i(0)} = 1.5,\ 2.25,\ 3$, and it explains the difference in the initial values of $\dfrac{Q_i(0)}{Q_{ci}}$ for the curves. 

As far as the general shape of these curves is concerned, it is similar to the classical resilience triangle as should be expected according the modern concept of the recovery processes. Namely, the initial horizontal fragment ends when the first resource quantum reaches the given city, the intermediate fragment exhibits the recovery to the minimal operating standards followed by the saturation meaning the finishing of the short-term recovery.

\begin{figure}
\begin{center}
\includegraphics[width=0.7\columnwidth]{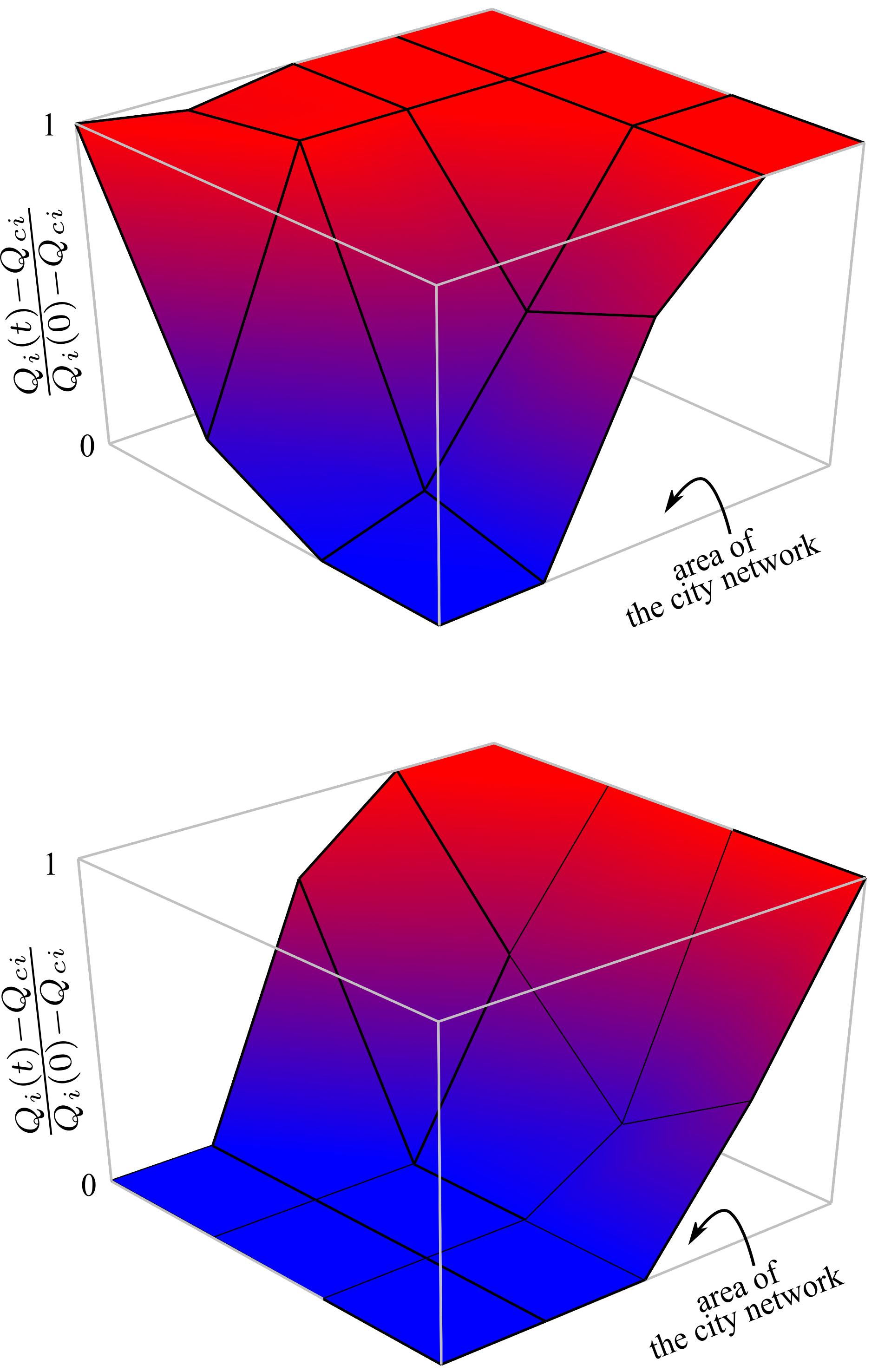}
\end{center}
\caption{The spatial pattern of the extra volume of resources distributed in the system after the recovery process has finished for two different degrees of damage, $a = 1.5$ (top) and $a = 3$ (bottom). As clearly shown, in the second case the number of cities involved in resource redistribution is considerably more than in the first case.}
\label{F:PatD123}
\end{figure}

The present result demonstrates a significant influence of the cooperative effects on the recovery dynamics. In fact, let us compare case~1 and case~3 (Fig.~\ref{F:D123}). The number of quanta required to recover city~$i$ in case~3 is four times bigger than that of case~1. However, the total duration of the redistribution process increases by less than twice. It is because that the greater damage is caused by the disaster, the more cities are involved in the recovery process. Figure~\ref{F:PatD123} justifies this conclusion depicting the spatial pattern of the extra volume of resources distributed in the system after the recovery process has finished.

\begin{figure}
\begin{center}
\includegraphics[width=0.9\columnwidth]{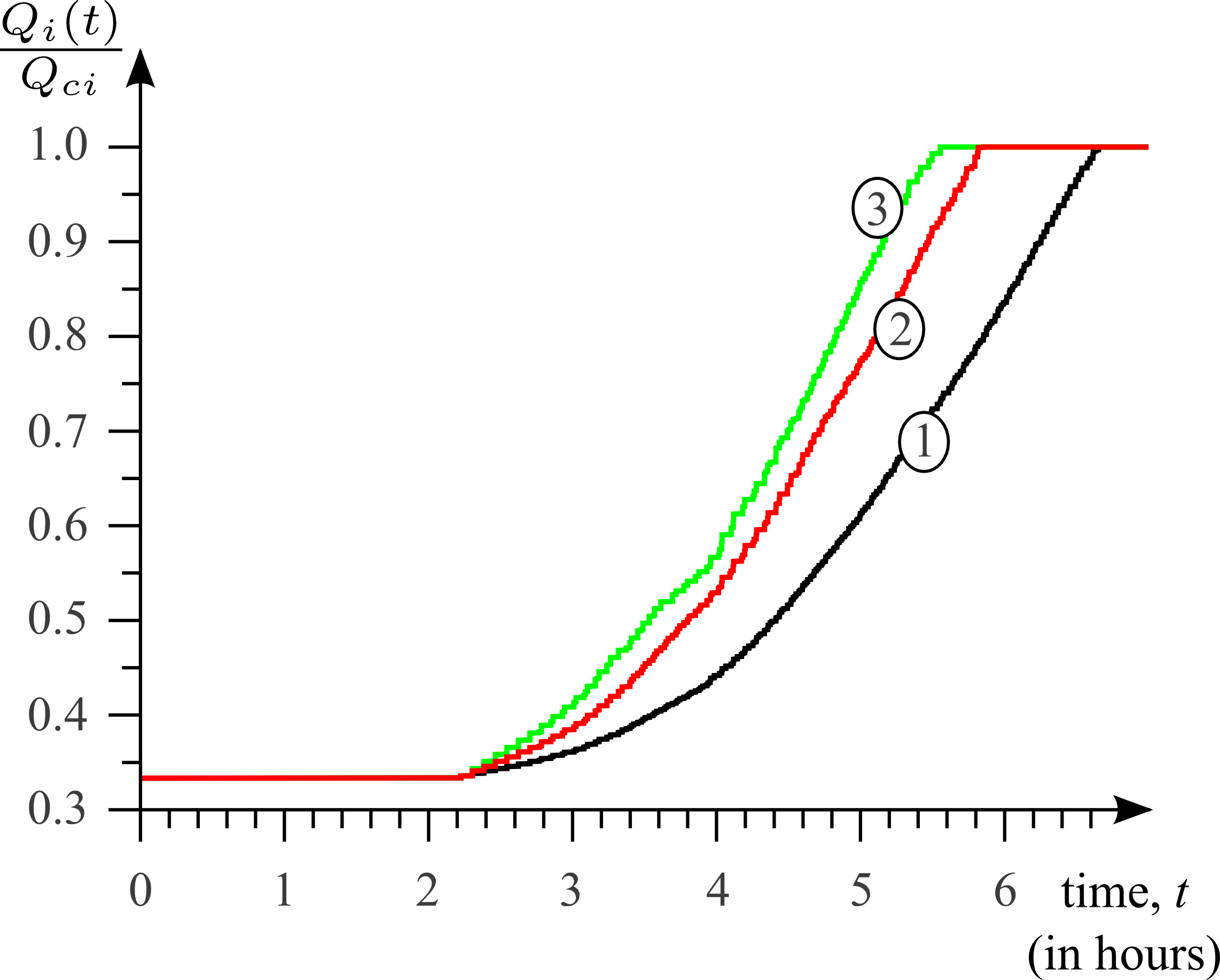}
\end{center}
\caption{The recovery dynamics of an affected city $i$. Curves 1, 2, 3 demonstrate the recovery dynamics when the capacity of cities  $c$ is equal to 15, 30, 45, respectively. All the other system parameters as well as the system topology were the same.}
\label{F:Q(t)C123}
\end{figure}

\begin{figure}
\begin{center}
\includegraphics[width=0.7\columnwidth]{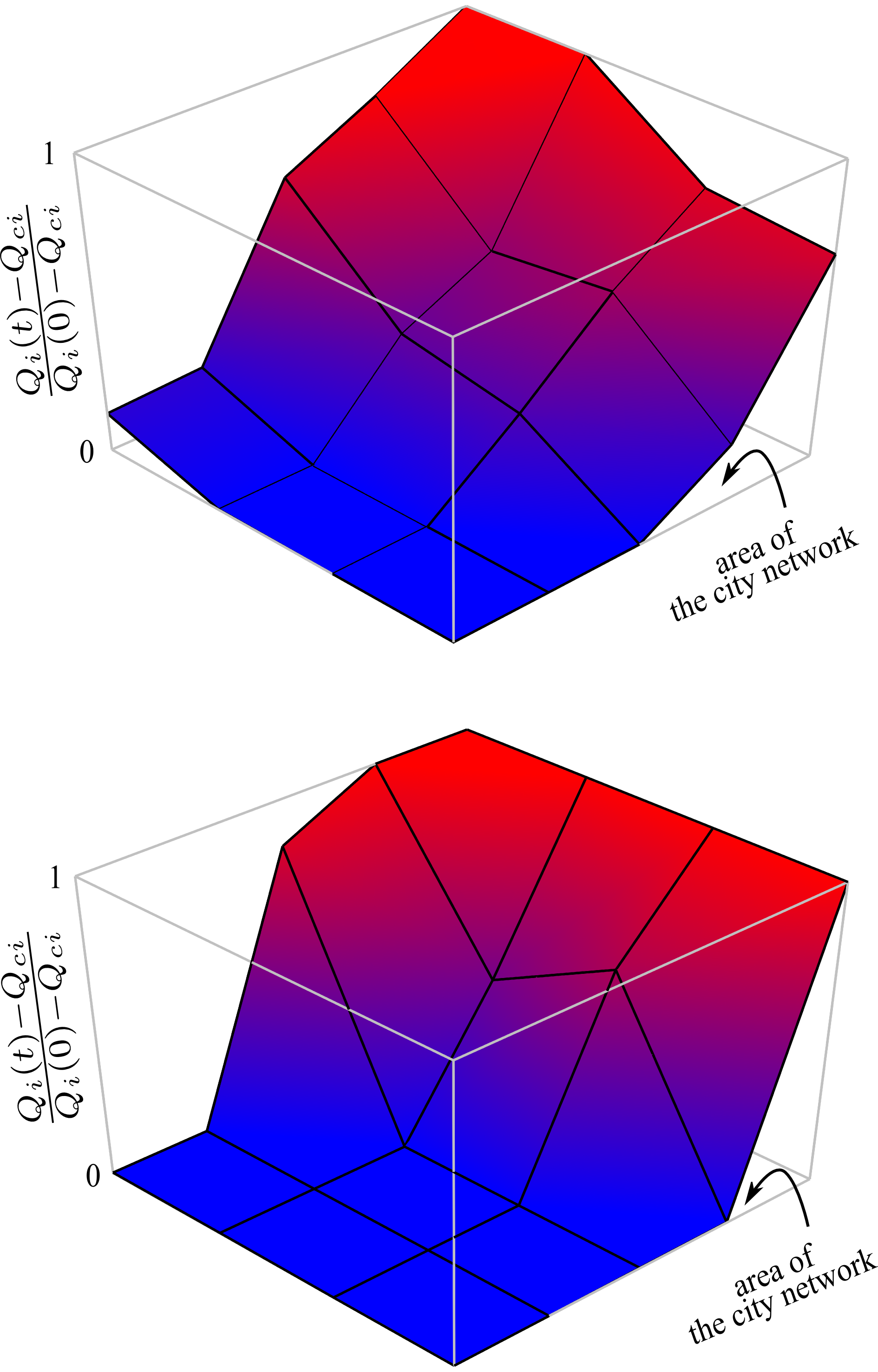}
\end{center}
\caption{The spatial pattern of the extra volume of resources distributed in the system after the recovery process has finished for two different values of the city capacity, $c = 15$ (top) and $c = 45$ (bottom). As clearly shown, in the second case the resource redistribution process is more local than in the first case.}
\label{F:PatC1C3}
\end{figure}

The next result is presented in Fig.~\ref{F:Q(t)C123} depicting the recovery dynamics of the affected city $i$ depending on the city capacity $c_i$ for a fixed damage degree, $a=3$. The capacity $c_i$ was changed from 15 to 45 for all the cities, which means that the number of quanta the cities are able to send per unit time was increased by three times in simulation. Nevertheless, the duration of recovery process changed only 1.5 hours (less than 30\% in relative units). It is also explained by the cooperative effects in the resource redistribution process, which is directly demonstrated in Fig.~\ref{F:PatC1C3}.

As shown in Fig.~\ref{F:PatD123}, both the quantities, the degree of damage $a$, and the city capacity $c_i$, affect the number of cities involved in the resource redistribution. One should, however, distinguish their effects. The degree of damage is not controllable parameter but a disaster characteristic, while the city capacity is a controllable technical parameter.

Some addition details of the effect caused by the city capacity are illustrated in Fig.~\ref{F:PatC1C3}. It depicts the spatial pattern of the extra volume of resources distributed in the system after the recovery process has finished for two different values of  the city capacity for the all cities, $c_i = 15$ (top) and $c_i = 45$ (bottom). We can see that in case of a smaller capacity the number of cities involved in the redistribution process is more than in case of a larger capacity. At the same time, the number of cities that send out all their extra volume of resources is more for the larger capacity. On one hand, therefore, the larger the capacity, the smaller the region comprising the cities involved in the recovery process, i.e., the higher the locality of this process. On the other hand, the smaller the capacity, the less the number of cities being in close to the minimal operation standards. Thereby the choice (if possible) of various values of $c_i$ can be determined for specific reasons.

\begin{figure}
\begin{center}
\includegraphics[width=0.7\columnwidth]{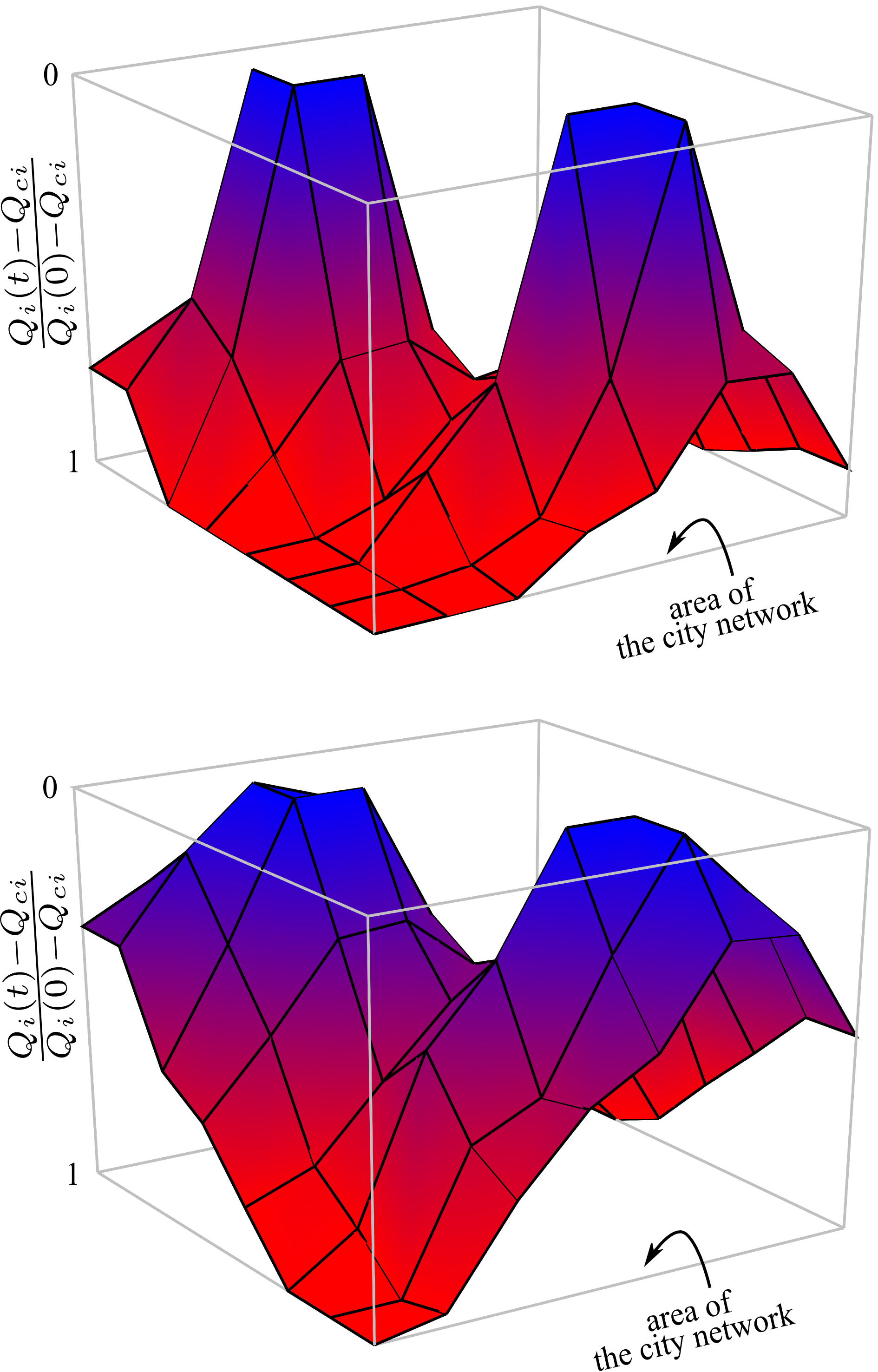}
\end{center}
\caption{The inverted spatial pattern of the extra volume of resources distributed in the system after the recovery process has finished when two separate regions interact directly. The top plot represents the spatial pattern for a relatively low degree of damage ($a = 1.5$); the bottom plot for a high degree of damage ($a = 3$).}
\label{F:Pat2AP}
\end{figure}

The effect of locality becomes more pronounced in the case where there are several separate groups of affected cities. To illustrate this we considered  a system with the increased number of cities from 20 to 64 and assumed that the affected cities belong to two groups located in the opposite sides of the system region. The number of residents and the total amount of resources were increased proportionally. Figure~\ref{F:Pat2AP} exhibits the spatial pattern of the extra volume of resources distributed in the system after the recovery process has finished. The top plot presents this patten for a relatively low degree of damage ($a = 1.5$). The bottom plot shows it for a high degree of damage ($a=3$). As shown here, in the former case we can identify two subsystems that do not interfere with each other in resource redistribution. As the degree of damage grows, the redistribution process drives these subsystems to cooperate and operate as a whole. The latter case exemplifies this effect. The cities located in the middle of the system region became involved in the redistribution of resources for the both affected groups of cities. It explains the saddle-shaped surface shown in the bottom plot. We call this effect  ``interference''.

\begin{figure}
\begin{center}
\includegraphics[width=0.9\columnwidth]{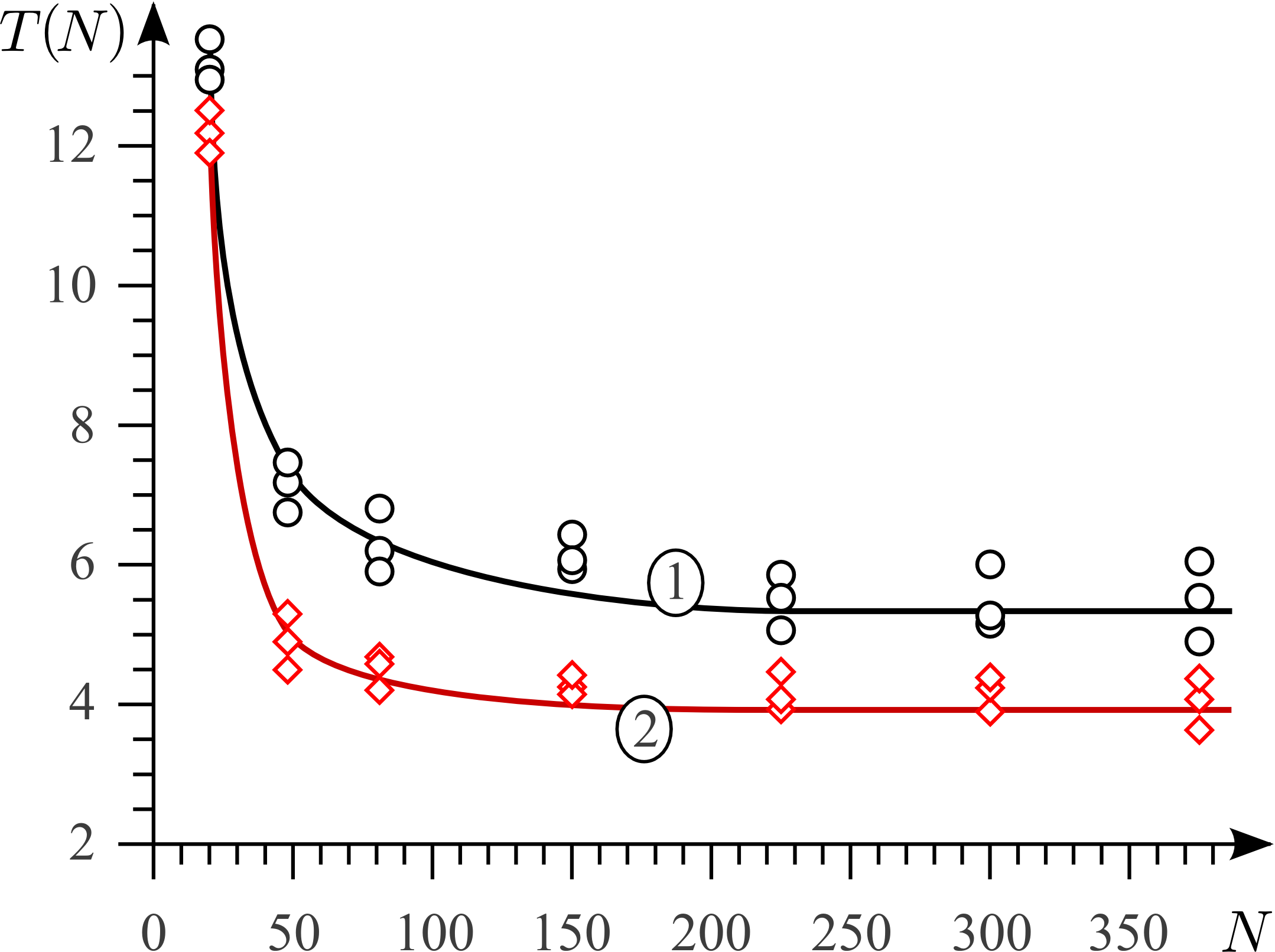}
\end{center}
\caption{The recovery duration $T(N)$  vs the number $N$ of cities that can be involved in the redistribution process. Curve~1 exemplifies the dependence for the case where the damaged cities are located in the corner, Curve~2 in the center of the system region. The dots represent the simulation results, and the curves are guides for the readers.}
\label{F:T(N)}
\end{figure}

The analyzed model uses a notion of administrative unit as an isolated system of cities individually responsible for the short-term recovery. Naturally, a cooperation of several administrative units can shorten the duration of this process. In order to study when such cooperation is efficient, we simulated the resource redistribution varying the number of cities (form 20 to 400) that can be involved in the process in principle. The result is presented in Fig.~\ref{F:T(N)} showing the dependence of the duration of redistribution process on the number of cities to be involved. The capacity of cities was set $15$, the time necessary to prepare one quantum of resources was increased by 3 times and set $0.25$ hour, and the degree of damage was set $4$. Curve~1 exemplifies this dependence for the case where the damaged cities are located in the corner, and Curve~2 in the center of the system region. For the given values of the system parameters the duration of the short-recovery exhibits fast drop within a interval from $20$ to $80$. It is explained by the fact that for this size of system all the cities are involved in the resource redistribution. When the system size exceeds some value around 100 cites, cities not participating in the process increase. When the distance between a given city and the affected region is far enough, it is more efficient to wait until a new quanta will be prepared in the neighboring cities than to request resources from the distance. It is responsible for the saturation in the dependence of the process duration vs the number of cities (Fig.\ref{F:T(N)}).

\subsection{``Centralized'' system}

Now let us consider the other type of system, i.e., the ``centralized'' one. In some sense it is the opposite type of the city network topology. Namely, we assumed that there are four big central cities of the equal size surrounded by 16 ``satellites'' (small cities), and 40\% of the system population are the residents of these centers. To compare the recovery dynamics for the ``centralized'' and ``uniform'' systems, the amount of critical resources was scaled with the number of residents such that the ratio $\dfrac{Q_{ci}}{N_i}$ is to be the same for the both types of systems.   
The amounts of resources in the ``satellites'' were set equal to their critical values $Q^\text{satellite}_i = Q^\text{satellite}_{ci}$, and all the ``extra'' resources of the system were concentrated in the centers such that $Q^\text{center}_i \gg Q^\text{center}_{ci}$. The total amount of resources and the system population were also equal for the ``centralized'' and ``uniform'' systems, i.e., Eqs.~\eqref{Eq.resource balance} and \eqref{Eq.residents balance}. The capacity $c_i$ of the centers was chosen twice as high as the city capacities for the ``uniform'' system, and the degree of damage was set $a = 2$. Besides, in order to smooth the discretization effects in the resource redistribution in the case under consideration we used the decreased volume of resource quantum, $h=0.2$, assuming it affords 20 residents.  

\begin{figure}
\begin{center}
\includegraphics[width=0.9\columnwidth]{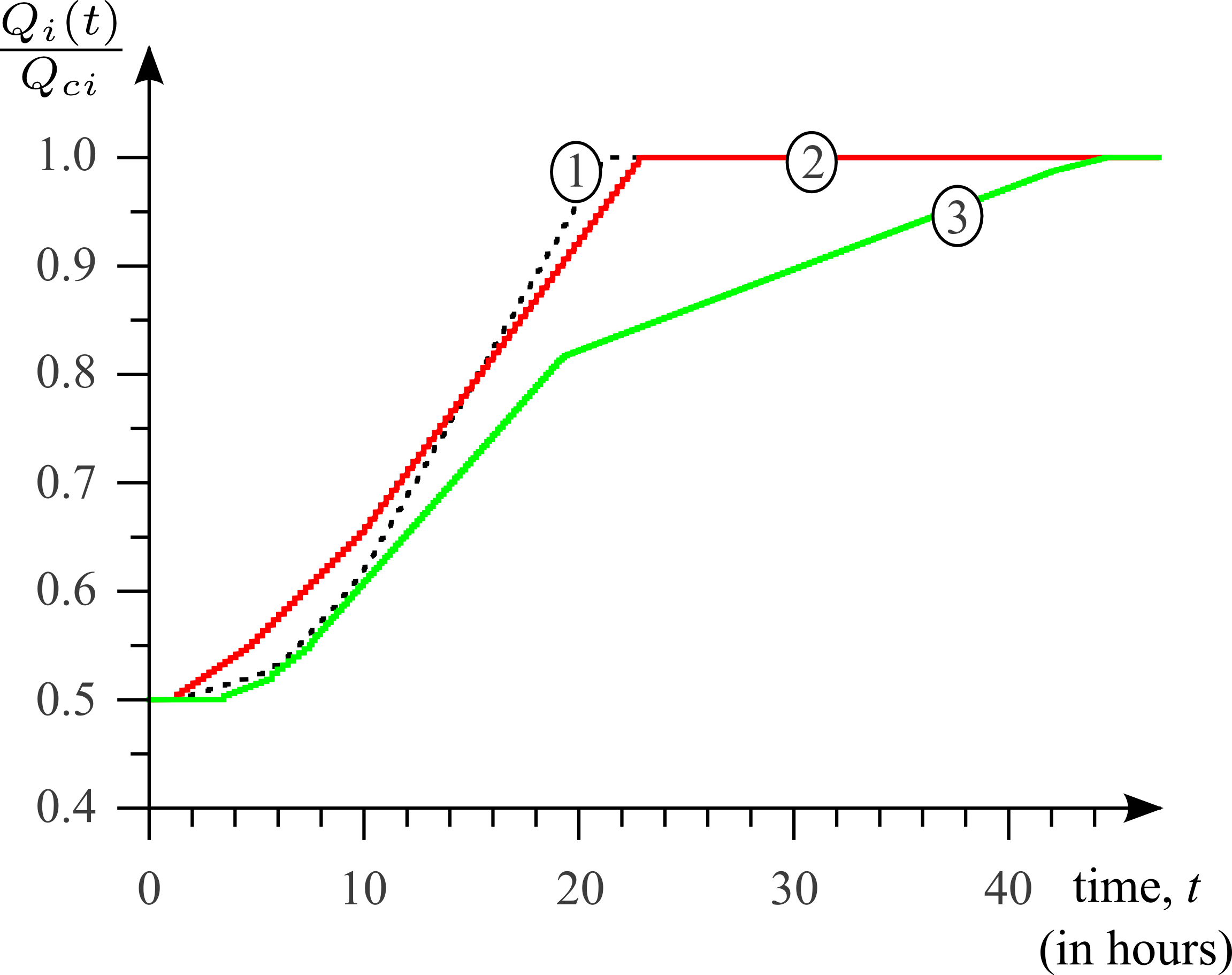}
\end{center}
\caption{The recovery dynamics of an affected city $i$ in three cases. Curve~1 demonstrates the recovery process in the ``uniform'' system, Curve~2 illustrates it when only ``satellites'' in the ``centralized'' system are damaged, and Curve~3 depicts the recovery dynamics of a damaged center.}
\label{F:Q(t)uc}
\end{figure}

Figure~\ref{F:Q(t)uc} compares the dynamics of the affected city $i$ in three cases. Curve~1 (dotted) shows the recovery dynamics for  the ``uniform'' system, Curve~2 that of a damaged ``satellite'' when all the centers are not affected. It should be pointed out that the duration of recovery process for the affected city in the ``uniform'' system turned out to be shorter than that of the ``satellite'', although the number of quanta requested by the ``satellite'' was 40~\% less than that for the city in the ``uniform'' system.  Curve~3 demonstrates that the recovery dynamics of the damaged center becomes twice as long as this process in the previous case. It is because the number of necessary quanta is much more and only the other three centers can be the donors of resources. Judging from the found results, the recovery process is more efficient for the ``uniform'' system than for the ``centralized'' one, if all the other factors being the same.

\begin{figure}
\begin{center}
\includegraphics[width=0.9\columnwidth]{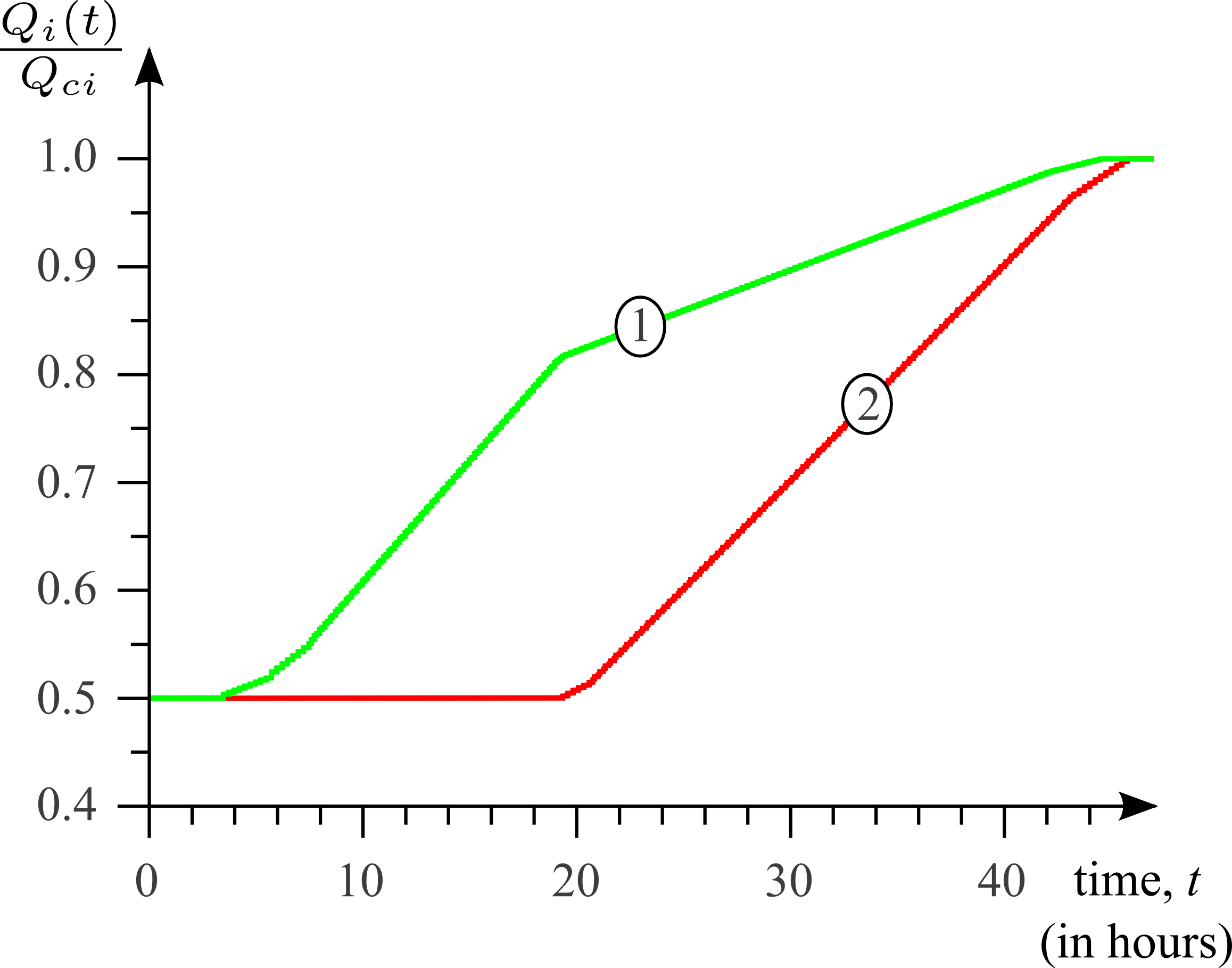}
\end{center}
\caption{The recovery dynamics in the case where the group of affected cities includes a center and ``satellites''. Curve~1 shows the recovery process of the center and Curve~2 the damaged ``satellite''.}
\label{F:Q(t)screen}
\end{figure}

The last Figure~\ref{F:Q(t)screen} illustrates an characteristic feature of the resource redistribution in the ``centralized'' system when the both types of cities (centers and satellites) are damaged. The number of residents in the damaged center as well as $Q_c$ are much larger than those in the ``satellite'', respectively. Therefore, the priority measure $S$ of the center is also higher. It explains that the resource flow from the donors is directed to the damaged center only for a relatively long time interval. Only when the priority measures of the center and ``satellite'' becomes equal, the resource flow is shared between them. We call it the ``screening'' effect.  

Figure~\ref{F:Q(t)screen} also demonstrates a general characteristic property of the resource redistribution governed by the developed algorithm. Even if the damaged cities are different in such parameters as $N_i$, $Q_i$, $Q_{ci}$, etc. the recovery process is completed practically at the same time.  It is one of the necessary properties for the algorithm to be strictly optimal. Therefore, we regard the constructed mechanism as semi-optimal. The question who it is close to the strictly optimal algorithm is worthy of individual analysis.

\section{Conclusion}

The short-term recovery of a region damaged by a large scale disaster has been under consideration. The short-term recovery  can be represented as a collection of actions with the common goal of restoring the corresponding life-support system to the minimal operating standards. The implementation of these actions could require a sufficiently large amount of resources (pure water, food, medical drugs, fuel, etc.) that are not available in the affected cities. Therefore, the resource redistribution of the vital resources becomes one of the key tasks of the short-term recovery.    

The present paper has proposed an algorithm by which one can construct a semi-optimal plan of implementing the resource redistribution. Its features are as follows. First, since this plan is created via a certain algorithm using the data collected after the outbreak of disaster, it is not based on any pre-planning. It is suitable, because the location, time, and consequences of the disaster are unpredictable. Second, the corresponding resource redistribution is a decentralized process in that there are no predetermined centers through which the main part of resource flow passes and is governed by it.  
%
Naturally the headquarter is responsible for the collection of information, its processing, and acceptance of the generated plan for implementation. Thereby we imply that the process implementation is decentralized whereas its management could be centralized.  
%
%
The decentralized resource redistribution enables the system to react to a disaster practically immediately and makes the recovery process cooperative. Due to the cooperative effects the size of the region involved in the recovery process becomes controllable.      

The proposed algorithm includes the following. Each city $i$ is characterized by the initial amount $Q_i$ of vital resources, its critical level $Q_{ci}$ and the number $N_i$ of residents. As a result of disaster, the critical level in the affected  cities is assumed to exceed the initial amount of resources, $Q_{ci}>Q_i$; in the other cities the opposite condition $Q_{ci}<Q_i$ holds.  The key point of the developed algorithm is how to deliver the required amount of vital resources to the affected cities from the neighboring ones in a certain semi-optimal way minimizing the duration of the recovery process. To measure the lack of resources in a given city, the quantity $\theta_i=\dfrac{Q_{i}-Q_{ci}}{Q_{ci}}$ has been introduces and the value $S_i=\theta_iN_i$ has been used to order the damaged city according to the priority of resources to be received. The cities are also characterized by the limit capacity of preparing and sending quanta of resources. Exactly this limit capacity endows the resource redistribution process with nonlinear properties. The matter is that when the limit capacity is attained, the ability of a city to send a new quantum is depressed for the time necessary to prepare it. The developed algorithm simulates this effect via temporal renormalization of the real time distances between the cities.  

The main attention has been focused on the recovery dynamics for the ``uniform'' system studied numerically. In particular, it has been demonstrated that a significant growth of the degree of damaged matches much weaker increase in the duration of recovery process. It is due to the cooperative effects in the resource redistribution, namely, the higher the damage degree, the more the number of cities involved in the resource delivery. Second, the city limit capacity is a controllable characteristic of the system that can affect the size of the region involved in the resource redistribution as well as the portion of these cities whose state drops to the minimal operation conditions after the recovery process has finished. 

The constructed model uses the notion of administrative unit that can implement the short-recovery process on its own. The conducted numerical simulation demonstrated that for each particular situation the dependence of recovery duration on the number of cities that can be involved in the resource redistribution exhibits saturation as this number increases. Actually it specifies the dimensions of the most appropriate administrative units that are to be involved in the disaster mitigation.

The recovery dynamics in the ``uniform'' and ``centralized'' systems has been compared. The latter system was assumed to contain just four big centers able to supply the surrounding ``satellites'' with vital resources. It has been demonstrated that in this case the cooperative effects are depressed giving rise to an increase in the recovery duration. If one of these centers is affected, the duration of the recovery process increases drastically. 

Besides, as found out in the case where the center and ``satellites'' are affected simultaneously, the individual recovery processes finish for all the cities practically at the same time in spite of the difference of the cities in size, population, and the required amount of vital resources. It is one of the necessary feature for an algorithm to be optimal and allows us to call the proposed recovery plan semi-optimal.











\bibliography{mylibrari}

\end{document}